\documentclass[a4paper,fleqn,usenatbib]{mnras}

\usepackage{newtxtext,newtxmath}

\usepackage[T1]{fontenc}
\usepackage{ae,aecompl}
\usepackage{pdflscape}

\usepackage{graphicx}	
\usepackage{amsmath}	

\newcommand{\oversim}[2]{\protect{\mbox{\lower0.5ex\vbox{%
  \baselineskip=0pt\lineskip=0.2ex
  \ialign{$\mathsurround=0pt #1\hfil##\hfil$\crcr#2\crcr\sim\crcr}}}}}

\title[Co-rotating structures around  NGC\,4490/85]
{A co-rotating gas and satellite structure around the interacting galaxy pair NGC\,4490/85}

\author[Karachentsev \& Kroupa]
{Igor D. Karachentsev$^{1}$\thanks{idkarach@gmail.com},
  Pavel Kroupa$^{2,3}$
\\
$^1$Special Astrophysical Observatory, The Russian Academy of Sciences, Nizhnij Arkhyz, Karachai-Cherkessian Republic
369167, Russia\\
$^2$Helmholtz-Institut f\"ur Strahlen- und Kernphysik, University
of Bonn, Nussallee 14-16, D-53115 Bonn, Germany\\
$^3$Charles University in Prague, Faculty of Mathematics and Physics,
Astronomical Institute, V Hole\v{s}ovi\v{c}k\'ach 2, CZ-180
00 Praha 8, Czech Republic\\
}

\pubyear{20xx}

\begin{document}
\label{firstpage}
\maketitle

\begin{abstract}
  The interacting binary system NGC\,4490/85 $=$ Arp\,269 is
  intermediate in mass between the Milky Way/Large Magellanic Cloud
  and the Large/Small Magellanic Cloud binary systems. It is a system
  of~14 known galaxies.  We estimate the total Newtonian gravitating
  mass of the NGC\,4490/85 group to be
  $M_T = (1.37\pm0.43) \times 10^{12} M_{\odot}$ using radial
  velocities and projected separations of its 13~candidate members.
  The system of dwarf satellites in the group demonstrates signs of
  coherent rotation in the same direction as that of the extended
  HI-shell surrounding the central interacting galaxy pair. The origin
  of this phase-space correlated population of star-forming late-type
  satellite galaxies raises questions in view of the
  planes-of-satellites observed around more massive galaxy pairs that
  are, however, made up of old early-type dwarf galaxies.  We also
  report the detection of a candidate stellar Plume near the binary.
  This elongated structure of low surface brightness is a likely
  optical counterpart to the HI-tail north of NGC\,4490/85, recently
  discovered by the FAST radio telescope.
  \end{abstract}

\begin{keywords}
  galaxies: dwarf; galaxies: groups: individual: NGC\,4490 group;
  Local Group; Magellanic Clouds; galaxies: kinematics and dynamics;
  cosmology: theory
\end{keywords}
 
\section{Introduction}
The discovery of planes of satellites around nearby luminous galaxies
has become one of the most important developments in 100~kpc-scale
cosmology. A significant portion of Milky Way's (MW's) dwarf
satellites are concentrated in a thin disk of satellites (DoS), the
members of which show signs of co-rotation
\citep{lyn1976,kro2005,metz2008,paw2020}.  A similar
planes-of-satellites structure has been found around the nearest
spiral galaxy M31 \citep{koch2006,metz2007,iba2013}.  The rotating
planar satellite structures of the Milky Way and of M31 appear to be
correlated with each other \citep{paw2013}. This has been interpreted
to be evidence for an encounter between the two galaxies about~8
to~10~Gyr ago \citep{bil2018, ban2022}.  Moreover, the evidence for
planar structures composed of dwarf satellites has been noted to also
exist around neighbouring galaxies: Centaurus-A
\citep{tul2015,mul2021}, M81 \citep{chi2013} and NGC\,253
\citep{mar2021}.  All these 5 galaxies have a high K-band luminosity
in the range of $(0.5 - 1.0)\times 10^{11}~L_{{\rm K}\odot}$ and are
located in a sphere with a radius of 4~Mpc around us.  In fact, every
single galaxy brighter than $L_K = 5\times 10^{10}~L_{{\rm K}\odot}$
in the nearest volume has been found to have a flattened
satellite-galaxy system.  The MW, M31 and NGC~253 systems have a ratio
of thickness to diameter of about~0.1, being seen nearly edge-on,
while the Cen~A and M81 systems are observed at an inclination. In
addition to the satellite-systems within a few~Mpc distance of us, a
survey of regions spanning~150~kpc around galaxies within a redshift
$z<0.05$ shows host galaxies to prefer, with 99.994~per cent
confidence, to have phase-space correlated satellite systems
\citep{iba2014, iba2015}. Additional flattened and disk-like
  satellite systems have been reported \citep{hees2021, cros2023}.

Here we present evidence for the presence of co-moving
satellites around the spiral galaxy NGC\,4490 with a K-band luminosity
of $1.9\times 10^{10}~L_{{\rm K}\odot}$.

The paper is organised as follows: Section~\ref{sec:properties} presents
the basic properties of the~14 dwarf galaxies forming a group together
with NGC\,4490. Section~\ref{sec:system} presents some
peculiar features of the group. Section~\ref{sec:data} reports the
discovery of a new candidate stellar structure in the vicinity of the
NGC\,4490/85, which we call the ``Plume''.  Section~\ref{sec:concs}
contains a discussion and the conclusions.

 \section{Properties of the NGC\,4490/85 galaxy group}
 \label{sec:properties}

 The spiral galaxy NGC\,4490 together with its neighbour, NGC\,4485,
 form the known pair Holm\,414 = Arp\,269 = VV\,30 = KPG\,341. The
 galaxies NGC\,4490 and NGC\,4485 have, respectively, radial
 velocities of~623~km~s$^{-1}$ and 517~km~s$^{-1}$ in the Local Group
 rest frame.  According to the Extragalactic Distance Database
 ($=$EDD, \citealt{ana2021}), the pair is located at a distance of
 8.91~Mpc, which we use in our calculations. Thus $1'=2.59\,$kpc at
 that distance. NGC\,4490 and its in-contact-neighbour, NGC\,4485,
 have K-band luminosities of, respectively,
 ${\rm log}_{10}\left(L_{\rm K}/L_{{\rm K} \odot}\right) = 10.28$
 and~8.99 (Table~\ref{tab:galaxies} below). For comparison, the
 corresponding log-luminosities of M31, the MW, the Large Magellanic
 Cloud (LMC) and the Small Magellanic Cloud (SMC) are, respectively,
 10.73, 10.78, 9.42 and~8.85 \citep{kar2013a}\footnote{For the
   up-to-date Local Volume database see www.sao.ru/lv/lvgdb/
   \,.}. Thus, given their luminosities, the NGC\,4490/85 pair is
 intermediate between the MW/LMC and the LMC/SMC interacting binaries.
 The projected on-sky distance between NGC\,4490 and NGC\,4485 is
 8.8~kpc.

 The group of galaxies around NGC\,4490 has in total currently-known
 15~proposed members as listed in Table~\ref{tab:group}. The columns
 of Table~\ref{tab:group} contain: (1)~--- the name of the galaxy:
 (2)~--- equatorial coordinates in degrees; (3)~--- super-galactic
 coordinates; (4)~--- morphological type in de Vaucouleurs digital
 scale; (5)~--- distance to the galaxy; (6)~--- the method by which
 the distance is estimated: via the tip of the red giant branch
 (trgb), by the Tully-Fisher relation (TF), by radial velocity of the
 galaxy in the Numerical Action Method (NAM) \citep{sha2017}, by
 probable membership in the group (mem); (7)~--- radial velocity of
 the galaxy relative to the Local Group centroid; (8)~--- luminosity
 of the galaxy in the $K$- band; (9)~--- projected separation of the
 galaxy from NGC\,4490; (10)~--- radial velocity of the galaxy
 relative to NGC\,4490; (11)~--- orbital estimate of the total
 Newtonian gravitating mass of the group via each satellite,
\begin{equation}
M_T = (16/\pi)\times G^{-1} \Delta V^2~R_p \, ,
\label{eq:MT}
\end{equation}
where $G$ is the gravitational constant.  The estimate of $M_T$
assumes an isotropic distribution of orbits of satellites with an
average eccentricity $\langle e^2\rangle = 0.5$ \citep{kar2021}.  The
values given in Table~\ref{tab:group} are taken from the latest version of the
Updated Nearby Galaxy Catalog (=UNGC) \citep{kar2013a} available at
www.sao.ru/lv/lvgdb.  Links to individual distance estimates are
provided in the last column of the table and notes to it.

    \begin{table*}
      \caption{Properties of the NGC\,4490/85 group members.
        The radial velocity errors are in the range of [1 -- 17]~km/s.
The two dwarfs without velocity estimates have a low surface
brightness and will need to be observed in the future.
}
      \label{tab:group}
\begin{tabular}{lccrclcrrrrc}\hline

 Name        &   RA (2000.0) DEC & SGL  SGB & T &  $D$  & meth& $V_{LG}$ &  $\log_{10}(L_k)$  & $R_p$ & $\Delta V$ &$M_T$ &    Ref.
\\
        \hline
            &     deg      &      deg    &    &  Mpc &   &km~s$^{-1}$
                                                                         &$L_{\rm K{\odot}}$  & kpc &  km~s$^{-1}$  &   $10^{11}~M_{\odot}$ & \\ \hline

   (1)      &     (2)      &     (3)  &    (4) &  (5) &  (6)  &   (7)   &  (8) &  (9) &   (10)  &   (11) &    (12)
\\ \hline

Dw1224+39   & 186.144+39.636 &   76.40+04.27 &    10  &  8.91 &   mem  &    ---  &   6.59 &   358  &   --- &   ---  &     ---
 \\
DDO 129     & 187.184+43.224 &   73.16+06.04 &     8  &  8.90 &   TF   &    582  &   8.81 &   249  &   --41&    4.9  &     [1]
\\
KK 149      & 187.218+42.178 &   74.18+05.77 &    10  &  8.51 &   trgb &    450  &   8.15 &    97  & --173&   34.2  &     [2]
 \\
Dw1229+41   & 187.430+41.162 &   75.21+05.65 &   --2  &  8.91 &   mem  &    ---  &   7.17 &    79  &   ---&    ---  &   --- 
  \\
Plume      & 187.555+42.017 &   74.41+05.97 &    10  &  8.91 &   mem  &    577  &   8.07 &    59  &   --46&    1.5  &   ---
  \\
KK 151      & 187.559+42.901 &   73.56+06.24 &     9  &  8.20 &   trgb &    479  &   7.79 &   196  &  --144&   48.0  &    [3]
 \\
NGC 4485    & 187.630+41.700 &   74.73+05.94 &     8  &  8.91 &   trgb &    517  &   8.99 &     9  &  --106&    1.2  &    [2]
 \\
NGC 4490    & 187.652+41.644 &   74.79+05.94 &     7  &  8.91 &   mem  &    623  &  10.28 &     0  &     0&    ---  &    ---
 \\
MAPS1231+42    & 187.788+42.094 &   74.38+06.16 &    10  &  8.13 &   trgb &    593  &   7.06 &    72  &   --30&    0.8  &    [4]
\\
UGC 7678    & 188.002+39.832 &   76.61+05.70 &     9  &  9.08 &   trgb &    710  &   8.24 &   285  &    87&   25.4  &    [2]
\\
UGC 7690    & 188.112+42.704 &   73.85+06.55 &     8  &  8.91 &   mem  &    578  &   8.66 &   173  &   --45&    4.1  &    ---
 \\
PGC 41749   & 188.470+39.626 &   76.91+05.99 &     9  &  9.94 &   trgb &    677  &   7.67 &   328  &    54&   11.3  &    [2]
 \\
UGC 7719    & 188.502+39.019 &   77.50+05.85 &     8  &  9.20 &   TF   &    704  &   8.21 &   420  &    81&   32.5  &    [1]
\\
Dw1234+41   & 188.659+41.276 &   75.35+06.57 &     9  &  8.45 &   NAM  &    638  &   7.26 &   130  &    15&    0.3  &    [5]
\\
UGC 7751    & 188.799+41.061 &   75.58+06.61 &     9  &  7.90 &   TF   &    641  &   7.49 &   161  &    18&    0.6  &    [1]
 \\
\hline
 \multicolumn{12}{l}{{\bf Notes}: [1]  \citet{kar2013a}, [2] \citet{ana2021}, [3] \citet{kar2018},}\\
 \multicolumn{12}{l}{[4]  \citet{tik2022}, [5] \citet{sha2017}} \\
\end{tabular}
\label{tab:galaxies}
\end{table*}

It should be noted here that for the dominant member of the group,
NGC\,4490, the EDD database \citep{ana2021} gives a trgb-distance of
7.07~Mpc (rather than~8.91~Mpc which is the distance of NGC\,4485
adopted here).  In this case, the components of the interacting pair
NGC\,4490/85 would be separated by a 3D distance of 1.84~Mpc, which
would question them being a physical pair and would disfavour a
relatively recent intense interaction of the galaxies. Analysis of the
color--magnitude diagram presented in EDD shows that the position of
the trgb on it is determined unreliably. Therefore, we adopted a
single value of 8.91~Mpc for both components of the pair.

Relative to NGC\,4490, the 12 companions (Table~\ref{tab:group}) with
measured radial velocities have an average velocity of
$\langle\Delta V\rangle = -27\pm26$~km~s$^{-1}$. The average distance
of 5 companions with accurately measured trgb-distances is
$8.80\pm0.27$~Mpc, in agreement with the adopted distance of
8.91~Mpc for the central galaxy pair.  This can be regarded as
evidence that the galaxies in Table~\ref{tab:group} (except
Dw\,1224+39) are real members of a single group.

\citet{elm1998} took deep images of the NGC4490/85 pair with the 0.6-m
Burrell telescope and noted several tidal features around the object
and star-forming regions.  \cite{huch1980} and \cite{cle1998}
investigated this interacting pair in the HI 21-cm line using the
100-m Effelsberg radio telescope and the Very Large Array (VLA).  They
found that the pair is immersed in a common prolate HI-shell, the size
of which reaches 60~kpc.  The last authors suggested that the
distribution of neutral hydrogen can best be explained by a
galactic-scale bipolar outflow of HI driven by supernovae in
NGC\,4490. The recent deeper HI observations with the
Five-Hundred-meter Aperture Spherical radio Telescope (FAST)
demonstrate that the HI-envelope has an extension of more than
100\,kpc \citep{liu2023}.  At the northern end of the HI-shell, the
authors found an HI-tail that stretches in the direction of the nearby
dwarf galaxy KK\,149.

\begin{figure*}
  \begin{center}
   \includegraphics[height=24cm]{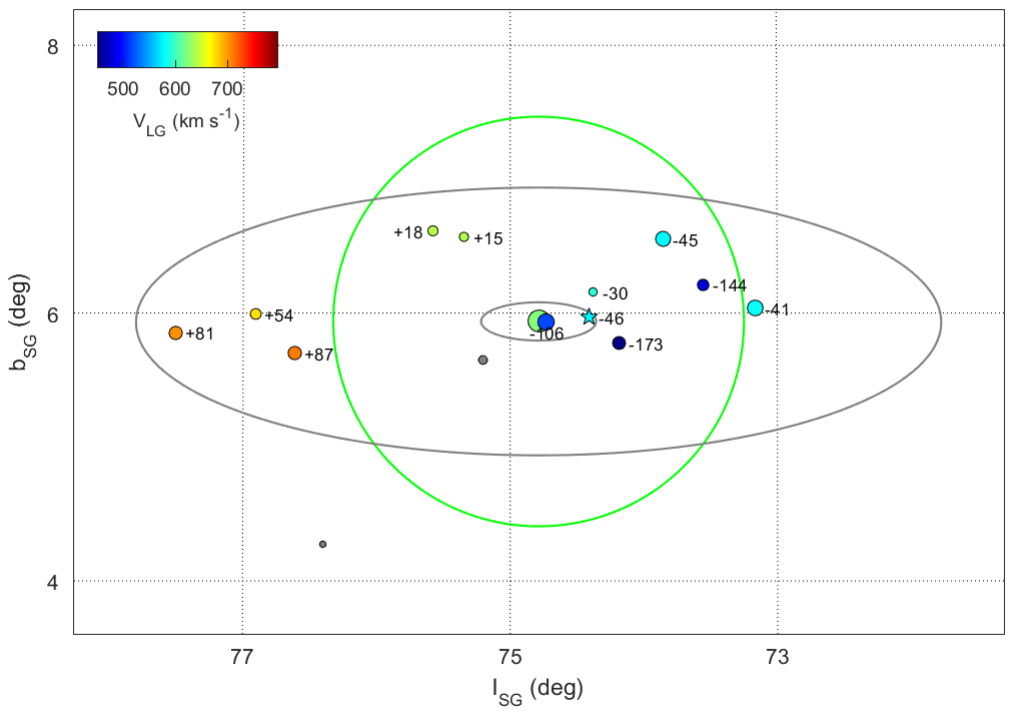}
\vspace{-7cm}
\caption{A view of NGC\,4490 galaxy group in super-galactic
  coordinates.  Dwarf companions of NGC\,4490 with positive and
  negative radial velocities relative to the main galaxy are marked by
  coloured circles according to a velocity scale in the left upper
  corner. The circles size corresponds to the galaxy luminosity. The
  asterisk shows the position of the new stellar structure, the Plume
  (see Sec.~\ref{sec:data}). The small central ellipse indicates the
  HI-shell around the interacting pair NGC\,4490/85. The large ellipse
  encloses the main body of the group.  The virial radius of the group
  is shown by the green circle and corresponds to 239~kpc.}
\label{fig:group}
\end{center}
\end{figure*}

The distribution of the 15~candidate members of the NGC\,4490/85 group
is presented in Fig.~\ref{fig:group} with circles, the sizes of which
correspond to the luminosity of the galaxy with color indicating the
galaxy radial velocity according to the scale in the left upper
corner. The numbers near the galaxies correspond to radial velocity
differences in km~s$^{-1}$ relative to the central galaxy NGC\,4490.
The new dwarf system, Plume (see Sec.~\ref{sec:data}), is shown by the
asterisk.  Two dwarf galaxies without measured velocities are marked
by grey circles. The small ellipse in the center of
Fig.~\ref{fig:group} reproduces the size and orientation of the
HI-shell around the interacting pair NGC\,4490/85 according to the
data by \citet{liu2023}.  The large ellipse encloses the main body of
the group.

The group of galaxies around NGC\,4490 has the following parameters:
an average projected separation of satellites
$\langle R_p\rangle = 182$~kpc, a mean-square radial velocity of
satellites $\sigma_v = 85$~km~s$^{-1}$, the average estimate of the
total Newtonian gravitating mass
$\langle M_T\rangle = (1.37\pm0.43) \times 10^{12}~M_{\odot}$.
According to \citet{tully2015}, the total (virial) Newtonian mass of a
group is related to the virial radius as
\begin{equation}
(R_v/ 215\,\,{\rm kpc}) = (M_T/ 10^{12} M_{\odot})^{1/3} \, .
\label{eq:Rv}
\end{equation}
This gives for the NGC\,4490 group the value $R_v = 239$~kpc, marked
with a green circle in Fig.~\ref{fig:group}.  With a total luminosity
of the group members
$\Sigma L_K = 2.20 \times 10^{10}\,L_{{\rm K}\odot}$, the total
Newtonian mass-to-luminosity ratio of the group is
$\langle M_T\rangle /\Sigma L_K = (62\pm20) M_{\odot}/L_{{\rm
    K}\odot}$.  This value is about half the global ratio of the mean
cosmic density of matter in the standard $\Lambda$CDM cosmological
{\it model}, $4.38 \times 10^{10} M_{\odot}/$~Mpc$^{^3}$ with
$\Omega_m = 0.3$, to the {\it observed} mean density of $K$-band
luminosity, $(4.3\pm0.2) \times 10^8~L_{{\rm K}\odot}$/Mpc$^3$
(table~4 in \citealt{driv2012}), equal to
$(102\pm5)\,M_{\odot}/L_{{\rm K}\odot}$.  Note that with
$M_T \approx 1.4 \times 10^{12}\,M_\odot$ within $R \approx 240\,$kpc,
the dynamical acceleration is $g \approx 0.03 \, a_0$, where
$a_0\approx 3.9\,$pc/Myr$^2$ is Milgrom's critical acceleration. The
Milgromian gravitating mass thus comes out to be
$M_M=(0.03 \, R)^2 \, a_0/G \approx 4 \times 10^{10}\,M_\odot$,
i.e. $M_M=(g/a0) \, M_T \approx 0.03\,M_T$, such that the K-band mass
to light ratio of the system is about~$2\,M_\odot/L_{\rm K \odot}$ in
Milgromian dynamics \citep{mil1983}.

Away from the main body of the NGC\,4490 group there is a dwarf galaxy
Dw\,1224+39.  It is located 15$\arcmin$ North of the spiral galaxy
NGC\,4369 $=$ Mrk\,439 that has a radial velocity of
$V_{LG} = 1038$~km~s$^{-1}$ and a TF-distance of 29.7~Mpc, and is
possibly a satellite of NGC\,4369.

The group in question is located near the Local Supercluster equator,
where the number density of galaxies at the line of sight is rather
high. Nevertheless, the NGC\,4490 group appears to be quite isolated.
There are no other galaxies with radial velocities in the range
(450--750)~km~s$^{-1}$ in the indicated area of size,
$7\degr\times4\degr$. According to \citet{kar2013b}, the spatially
closest neighbour to the NGC\,4490 group is the group of~18 galaxies
around NGC\,4258 ($12^h19^m +47\degr18\arcmin$) at a trgb-distance of
7.66~Mpc from the observer. The spatial separation between the group
centers is 1.56~Mpc.

\section{The system of satellites around NGC\,4490}
\label{sec:system}

The group of galaxies around the interacting pair NGC\,4490/85 is
characterised by the following features.

i. The group can be classified as a ``fossil'' group because the
luminosity of each satellite does not exceed 1/20th of the luminosity
of the principal galaxy. This difference in luminosities will
intensify if the components of the central pair were to merge.
This is expected to occur on a time scale of $\approx 5 \times 10^8$ years,
provided both are immersed in dark matter halos (but see
\citealt{kro2015, ros2021, has2022, kro2023}).

ii.  Almost the entire population of the group consists of late-type
galaxies with ongoing star formation. Only one dwarf galaxy,
Dw\,1229+41, near the interacting pair, is classified by us as a quiescent
dSph object.  Within the Fig.~\ref{fig:group} area, we did not find
other dSph galaxies with a luminosity brighter than
$L_K = 3\times 10^6\,L_{{\rm K}\odot}$.

iii.  The satellite system is stretched along the super-galactic
plane. In the equatorial coordinates, the positional angle of its
major axis (the large ellipse) is PA$= 158\degr\pm3\degr$.  The
positional angle of the HI-shell around the central interacting pair
(small ellipse in Fig.~\ref{fig:group}) is practically oriented in the
same direction, PA$= 159\degr\pm2\degr$. It is noteworthy that the
major axis of the Plume (Sec.~\ref{sec:data}) has also a similar
direction, PA$= 155\degr\pm5\degr$.

iv.  The satellite system reveals a distinct asymmetry in the
distribution of radial velocities relative to the main
galaxy~NGC~4490. All the objects on the northern side of the group
(bluish circles on the right side of Fig.~\ref{fig:group}) have
negative relative velocities, whereas all the objects on the opposite
side (reddish circles) have positive ones. This suggests the system of
satellites to be arranged in an inclined similar rotationally
supported DoS structure as observed around the MW, M31 and Cen~A.

v.  Moreover, as is seen on VLA- and FAST-maps, the HI-shell around
NGC\,4490/85 is also characterised by an increase in radial velocity
from the northern side to the South, conforming with the satellite system.

The listed geometric and kinematic coincidences do not look
accidental. The observational data show the gas component of
the central part of the group and the system of satellites as a whole
to have mutual coherent motions.

\section{The Plume: the discovery of a new stellar structure near
  NGC\,4490}
\label{sec:data}

About a dozen dwarf galaxies with radial velocities close to the
velocity of the centroid of the NGC\,4490/85 pair were known in its
vicinity.  Based on data from a recent DESI Legacy Imaging Surveys of
the sky \citep{dey2019}, \citet{kar2022} performed a search for new
dwarf satellites of NGC\,4490/85 and found three objects of low
surface brightness that could be companions of the pair. The recent
publication \citep{liu2023} gave us a reason to repeat the search for
faint dwarfs around NGC\,4490 using the DESI Legacy Imaging Survey
data. As a result, we discovered an object of extremely low surface
brightness, which is located in the region of the northern HI-tail
indicated by \citet{liu2023} on their figures. The object has
dimensions of $\approx 4\arcmin\times 1\arcmin$ and is elongated in
the direction of NGC\,4485 and KK\,149.  It will be important to
quantify this object using other deep imaging surveys (such as CFIS-r
at the CFHT, \citealt{iba2017}). It is not an instrumental artefact,
but could in principle be MW cirrus. However, its location in the
direction of a HI tidal tail makes it a possible candidate new stellar
structure, a stellar plume (hereinafter refered to as the Plume),
associated with the NGC\,4490/85 system.  Judging by its structure,
the new object may have been a past dwarf galaxy, destroyed in the
pair's gravitational field and surrounded by tidal material, being
similar to the time-dependent models of such systems \citep{kro1997,
  cas2012}.

  \begin{figure*}
  \begin{center}
    \includegraphics[height=20cm]{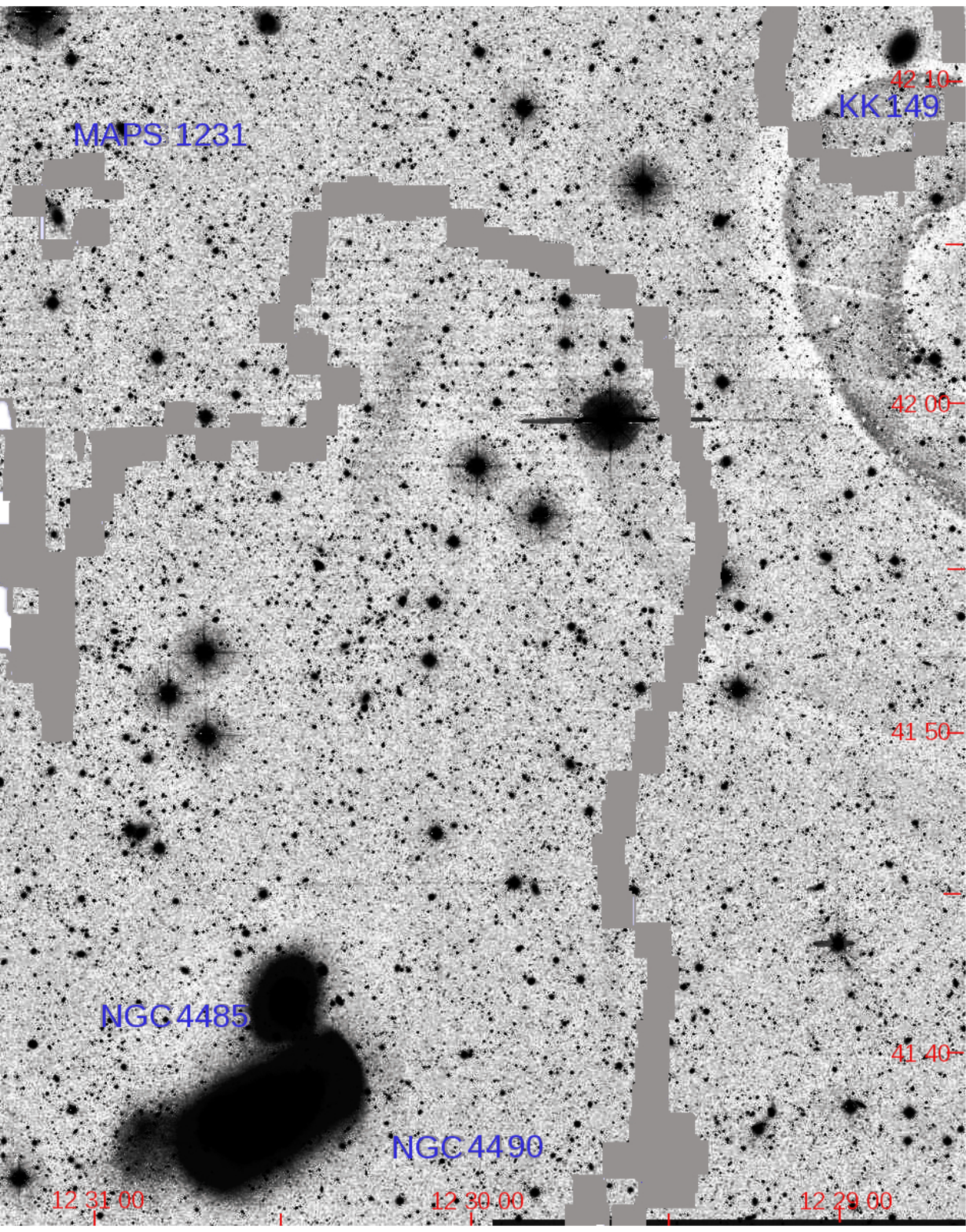}
    \caption{ The high-contrast image of the interacting galaxy pair
      NGC\,4490/85 and their dwarf satellites KK\,149 and MAPS1231+42
      from the DESI Legacy Imaging Survey.  The image size is
      $39\arcmin\times30\arcmin$, corresponding to $101 \times 78$~kpc
      at the adopted distance of 8.91~Mpc. North is up, East is left,
      red numbers indicating the right ascension and declination.  The
      sequence of grey squares represents the faintest (light-blue)
      contour of the HI-envelope around the interacting pair according
      to fig.~1 in \citet{liu2023}. Their size corresponds to the
      dimension of the HI- pixels. The elongated low-contrast Plume
      under consideration is located~8' to the North-North-East from
      the image center. The large oval structures on the upper right
      side of the figure are due to reflection from a bright star
      outside the image. }
\label{fig:Plume}
  \end{center}
\end{figure*}

\begin{center}
\begin{table}
\caption{Parameters of the Plume.
}
\begin{tabular}{l|c}\hline

  RA    (J2000.0)               &         $12^h30^m13.2^s$
\\
  DEC (J2000.0)                  &      $+42\degr01\arcmin01\arcsec$

\\
  Angular diameter              &          $4\farcm2$

\\
  Axial ratio                   &                 0.31

\\
  Positional angle, deg         &         155 

\\
%
                                        
\\
  
  HI-flux, Jy~km~s$^{-1}$              &  $\approx1.1$

\\
  $V_{\rm hel}$,~km~s$^{-1}$                   &           540
\\ \hline
  \end{tabular}
\label{tab:Plume}
\end{table}
 \end{center}
 The DESI Legacy Imaging Survey image of the object Plume can be found
 in Fig.~\ref{fig:Plume}.  The figure shows the interacting binary
 galaxies NGC\,4490 and NGC\,4485 together with their dwarf companions
 KK\,149 and MAPS1231+42 taken from the DESI Legacy Imaging Survey
 \citep{dey2019}.  The contour of the diffuse tail of HI recently
 mapped by the FAST radio telescope \citep{liu2023} is indicated.
 This apparently tidal system is barely seen in the Sloan Digital Sky
 Survey and very barely visible in the UV-survey GALEX. It is
 extremely difficult to determine the integral apparent magnitude of
 the Plume. We estimated it by the average brightness contrast of the
 Plume above the night sky background. With a contrast of about 1\%, a
 night sky brightness of 22.5 mag/sq.arcsec, and dimensions of
 $4.2\arcmin\times 1.2\arcmin$, the integral magnitude of the Plume is
 $B \approx 16\fm9$ with a central surface brightness in the $B$-band
 of $\approx27\fm5\,/\sq\arcsec$. The basic parameters of the Plume
 are presented in Table~\ref{tab:Plume}.  The magnitude of the HI-flux
 and the radial velocity of the Plume are estimated by us using the
 VLA- and FAST-maps presented by \citet{cle1998} and \citet{liu2023}.
 The HI-flux is very approximate due to unknown projection factors.
 Assuming a distance to the Plume of 8.91~Mpc, we roughly estimate its
 hydrogen and stellar masses as $2.0 \times 10^7~M_{\odot}$ and
 $8.0 \times 10^7~M_{\odot}$, respectively. The latter can be larger
 by an order of magnitude for a young system if the galaxy-wide IMF
 varies systematically as indicated by other observations
 \citep{jer2018}.

 The observational parameters of the Plume, given in
 Table~\ref{tab:Plume}, are poorly determined due to its very low
 surface brightness. Dedicated deep follow-up observations in
 different ($u,g,r,i$) filters will be needed to establish Plume's
 spectral energy distribution and to clarify whether it is young, as
 expected for a stellar population recently born from gaseous
 debris. The physical dimension of the Plume is
 $\approx 10\,{\rm kpc} \times 2.5\,$kpc if it is at the distance of
 the NGC\,4490/85 binary (Table~\ref{tab:Plume}). If this were to be
 the physical scale of this object, then it would be larger by a
 factor of ten than known tidal-dwarf galaxies of similar mass
 (c.f. fig.~2 in \citealt{dab2013, has2019}). It is thus likely
 unbound, consisting of an evolved version of the type of star-forming
 regions and stellar populations observed to be in tidal debris
 extended over dozens of~kpc in interacting galaxies. Examples of such
 tidal structures, plumes and streams are the putative tidal dwarf
 KK\,208 (PC166170) near NGC\,5236 with a Holmberg diameter of
 5.9~arcmin (8.8~kpc) \citep{kar2013a}, the eastern tidal structure of
 NGC\,5907 with a diameter of 12~arcmin \citep{lanz2023} corresponding
 to a linear dimension of 53~kpc at a distance of 14.3~Mpc, the
 western stream near NGC\,5907 being fainter and about twice as
 extended.  Other examples of such structures are the dog-leg tidal
 stream of NGC\,1097 \citep{gal2010}, tadpole galaxies
 \citep{elm2012}, the Leo Triplet \citep{wu2022} and other interacting
 galaxies \citep{rod2023} and the umbrella-like stellar stream of
 NGC\,922 \citep{mar2023a}. Such tidally-produced structures of
 low-surface brightness constrain the galaxy encounters that produced
 them which test cosmological structure formation theory
 (e.g. \citealt{kro2015}) and systematic surveys for them such as
 through The Stellar Stream Legacy Survey are underway
 \citep{mar2023b}. On the other hand, with a central surface
 brightness, $\approx 27.5\,$mag/square arcsec and an effective linear
 diameter of $\approx10\,$kpc, the Plume might also classify as being
 an ultra-diffuse galaxy (UDG, \citealt{vanD2015}), probably having
 been assembled during a past encounter between NGC\,4490 and
 NGC\,4485. We emphasise that the formation of UDGs remains not
 understood.
 
 The HI peaks at an RA of about 12:30:00 \citep{liu2023}, while the
 optical object is at 12:30:13, which is about 3' away.  In most
 interacting systems able to produce tidal dwarf galaxies (TDGs), the
 very young TDG coincides with the peak of the HI. More generally,
 with rare exceptions, the optical and HI tidal features overlap in
 space in such freshly formed dwarf satellite galaxies.  This is
 clearly not the case here, further pointing to the Plume being a
 stellar stream.  In this interpretation, the question arises why the
 southern HI tail (to the lower left but off Fig.~\ref{fig:Plume})
 which seems to have a similar column density as the northern one does
 not have an optical counterpart.

\section{Discussion and Conclusion}
\label{sec:concs}

The discovery of a hitherto not know stellar structure, the Plume, in
the NGC\,4490/85 group of galaxies is reported
(Sec.~\ref{sec:data}). The Plume might be part of a stellar stream or
a UDG possibly formed during a past encounter between NGC4490 and
NGC\,4485 (Sec.~\ref{sec:data}). The Plume needs dedicated follow-up
observations to improve understanding of its nature.

This work also reports the discovery that the NGC\,4490/85 group has a
rotating and thus highly phase-space correlated population of
satellite galaxies.  The kinematic pattern observed in the NGC\,4490
group indicates a joint co-rotation of the system of satellites and
the gas envelope surrounding the central interacting pair
NGC\,4490/85.  Assuming regular rotation, the satellite system will
have a significant angular momentum
$J_{\rm sat} = \Sigma~M_* R_p |\Delta V|$. We estimated $J_{\rm sat}$,
comparing it with the angular rotational momentum of the principal
galaxy, $J_{\rm N4490} = M_* V_m R_{26}$.  Here, the amplitude of the
galaxy rotation, $V_m = 86$~km~s$^{-1}$, and the galaxy optical
radius, $R_{26} = 8.1$~kpc, are taken from UNGC \citep{kar2013a}.
Assuming the same stellar mass-to-luminosity ratio for the central
galaxy and for its satellites, we obtained the value
$J_{\rm sat}/J_{\rm N4490} \approx 2$, which is approximate due to
uncertain projection factors in the motion of the satellites.  Thus,
the total angular momentum of the satellites turns out to be
comparable with the momentum of rotation of the dominant galaxy in the
group.

The satellite galaxies not only appear to lie in a rotating plane, but
also seem to be concentrated in a bar-like configuration, which is
slightly offset from the line $b_{\rm SG} = 6^{\rm o}.0$ (see
Fig.~\ref{fig:group}). Moreover, from the data in
Table~\ref{tab:group} the mass of the satellite system may be
concentrated on the east (left) side of this configuration. If this is
true, then it seems likely that the satellite system would exert a
tidal (and $m=1$) force on the outer HI distribution (which also seems
aligned along about the same line). Thus, another source of tidal HI
distortion could be the outer satellite system (see \citealt{cris2019}
for a discussion of such effects).

The dwarf satellites may be in primary infall towards the massive
central galaxy from both its sides along a putative diffuse cosmic
filament.  Such an assumption implies the possible presence of more
distant companions at distances $\approx (2 - 4) R_v$, which, in
principle, is accessible to observational verification. This
explanation however raises the question where all the primordial dwarf
satellite galaxies are that ought to be evident in the dark matter
halos of the NGC\,4490/85 galaxy pair according to the $\Lambda$CDM
model.

On the other hand, the system of phase-space correlated satellite
galaxies around the NGC\,4490/85 pair plus the co-rotating gas
envelope is reminiscent of the similar arrangement of MW satellite
galaxies and young halo globular star clusters in a vast polar
structure that co-aligns and co-orbits with the gaseous Magellanic
Stream \citep{paw2013, paw2020}.  The similarity of the angular
  momenta of the NGC\,4490/85 satellites and the dominant galaxy pair
  may be due to the former forming as a population of tidal dwarf
  galaxies from the latter. Such a process of the formation of new
  dwarf galaxies in gaseous tidal tails produced by an encounter of
  two late-type galaxies has been documented through simulations
  \citep{wet2007}. It has also been shown that the satellite
  distributions around the MW and of M31, on being correlated
  \citep{paw2013}, suggest their formation through a past encounter
  between the MW and M31 as shown by simulations \citep{bil2018,
    ban2022}. The NGC\,4490/85 satellite system differs from the other
  known phase-space-correlated satellite systems that are made up of
  old early-type dwarf-spheroidal galaxies, by being composed of
  late-type star-forming dwarfs. The satellites of NGC\,4490/85 would
  therefore be a freshly-formed system of tidal dwarf
  galaxies. Theoretical work via computer simulations is needed to
  study if the orbit of NGC4485 around NGC4490 can have expelled tidal
  tails that are sufficiently massive to form the observed population
  of satellite galaxies sufficiently recently to allow them to be
  still star-forming. The observational finding that even the
  relatively low-mass interacting binary system NGC\,4490/85 can have
  an associated phase-space correlated satellite systems composed of
  star-forming dwarf galaxies thus poses an interesting problem to
  understand theoretically. It might be related to the distribution of
  (the non-satellite) star-forming dwarf galaxies in a
  highly-symmetrical structure around the MW/M31 pair in the
  Local-Group \citep{paw2013}, a structure that has not yet found a
  theoretical understanding.

  While the individual plane-of-satellite system or DoS around the MW
  is being understood to not be a challenge by it being ``common (at
  the 1/231 level or at the 1~per~50~Mpc cubed level)''
  (\citealt{xu2023}, see also \citealt{saw2023}), according to
  \cite{ase2022} the MW, M31 and Cen~A phase-space correlated
  satellite systems are, combined, in~$5.27\,\sigma$ tension with the
  $\Lambda$CDM model.  This is confirmed by \cite{kan+2023} who
    show mergers cannot produce correlated systems of pre-merger
    satellite galaxies. Correlated satellite systems are formed from
    tidal-dwarf galaxies though, with \cite{has2019} therewith
    confirming the dual dwarf galaxy theorem for the $\Lambda$CDM
    structure-formation model. The real tension would be higher if
  the mutual correlation of the MW and M31 systems were taken into
  account.  The additional two systems (M81, NGC\,253) together with
  the present discovery of another highly phase-space correlated
  satellite plus circum-galactic gas system (NGC\,4490/85), imply
  the planes of satellite problem to be an until now not solved
  failure of the current
  cosmological structure formation theory \citep{kro2005, paw2018,
    paw2021a, paw2021b,kro2023}.

\section*{Data Availability}
The data on which this work is based is publicly available and is
detailed in the corresponding section of this manuscript.
 
\section*{Acknowledgements}

We thank the referee and Curtis Struck, Mordehai Milgrom, Marcel
Pawlowski and Oliver M{\"u}ller for constructive and helpful comments.
I.D. is grateful to Ming Zhu, Serafim Kaisin and Dmitry Makarov for
useful discussions.  This work has made use of DESI Legacy Imaging
Survey data, and the revised version of the Local Volume galaxy
database. The Local Volume galaxies database has been updated within
the framework of grant 075--15--2022--262 (13.MNPMU.21.0003) of the
Ministry of Science and Higher Education of the Russian Federation.


\end{document}